# Transformation of Networks through Cognitive Approaches

**[1]T.R.Gopalakrishnan Nair [2]Abhijith [3]Kavitha Sooda**

[1]Director- Research & Industry, trgnair@ieee.org
[2,3] Member, Research Group – Cognitive Networks – RIIC
abhijiths85@gmail.com, kavithasooda@gmail.com
[1,2,3] Dayananda Sagar Institutions, Bangalore, India

**Abstract**

The growth in data traffic and the increased demand for quality of service had generated a large demand for network systems to be more efficient. The introduction of improved routing systems to meet the increasing demand and varied protocols to accommodate various scales of challenges in network efficiency had further complicated the operations. This means that a better mode of intelligence has to be infused into networking for smoother operations and better autonomic features. Cognitive networks are defined and analyzed in this angle. They are identified to have the potential to deal with the future user related quality and efficiency of service at optimized levels. The cognitive elements of a system like perception, learning, planning, reasoning and decision forming can enable the systems to be more aware of their environment and offer better services. These approaches are expected to transform the mode of operation of future networks.

## I. Introduction

The recent growth in user demands and technology transformations in networking has lead to a paradigm shift in designing new modalities of operations of network with increased capabilities and ease of use [35]. The need of the hour is to ensure that the network adapts its behavior to the changes in networks, learns from its environment and exploits the available knowledge to improve its future behavior. This method over time has evolved into, a progressive design concepts called the cognitive network. This is an intelligent form of networks that is expected to meet the end-to-end goals of message transportation in a better way. This technology involves a number of intelligent approaches that facilitate the sharing and reuse of knowledge in many components that govern the performance of a total network. These tasks include perception, acting and planning, learning, reasoning and decision-making [1].

Cognition can be compared to a mental process which includes the above mentioned aspects.. In this, *Perception* is to learn from the environment and understand the changes in environment. *Acting and Planning* evolves from understanding the changes in environment. *Learning* is to understand and grow in the environment. *Reasoning* is to analyze the reasons or the motivations of the changes. And *Decision Making* is to decide what to do according to the results of reasoning to achieve a predefined goal.

## II. Cognitive Network

### A. Cognitive Approach

Current data network technology often limits itself to changes and interaction with the network resulting in sub-optimal performance.

Limited in state, scope and response mechanisms, the network elements (consisting of nodes, protocol layers, policies and behaviors) are unable to make intelligent adaptations. In contrast, cognitive network is a network composed of elements that, through learning and reasoning, dynamically adapt to varying network conditions in order to optimize end-to-end performance. This has evolved to meet the requirements of the network as a whole, rather than the individual network components [2].

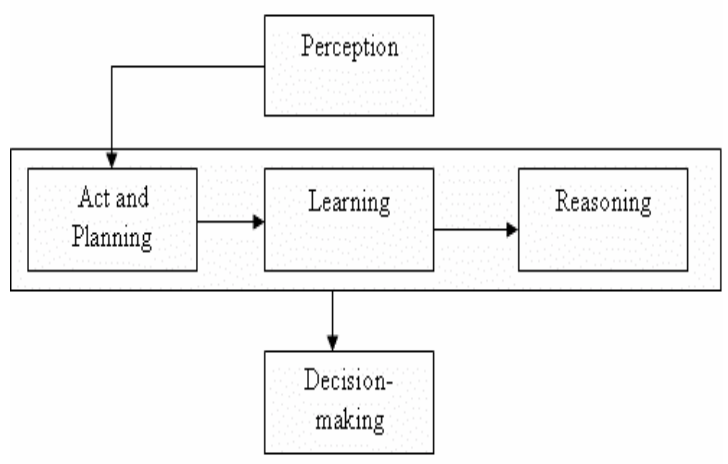

**Figure 1** Cognitive approach

Cognition was first defined by Mitola [3] and as defined by many others goes simply beyond adaptation. This definition converges to an idea of feedback loop [4]. This feedback loop was commonly called as OODA loop as shown in Figure 2. This is used by military officers for thought process of their opponents.

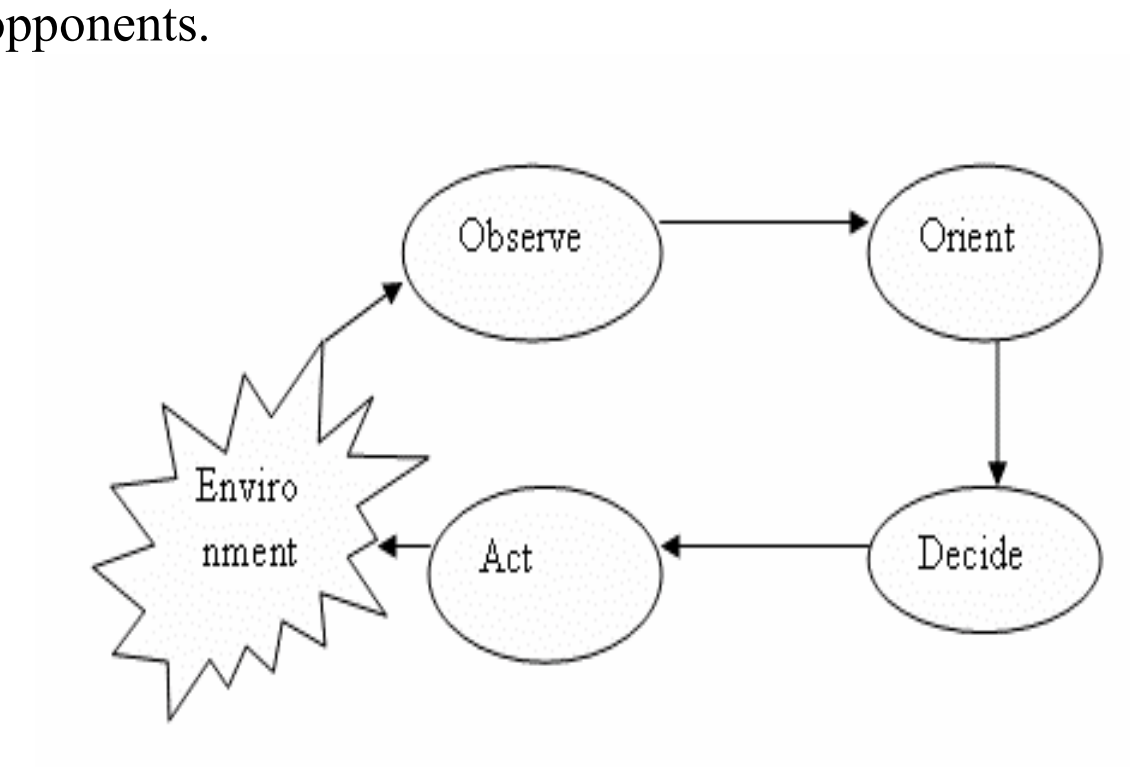

**Figure 2** OODA loop

Two important components need to be included in the OODA loop. One is the overarching module, which deals with the input to the loop and guides the orientation and decision components by providing a context in which to make a decision. The second one is the learning module, which prevents mistakes from previous iterations occurring in future iterations.

Most engineering problems deal with multiple metrics. In general, when a problem has multiple objectives it will not be able to optimize all metrics. This reaches a point where one metric cannot be optimized without affecting another. This area of optimization research is called Multiple Objective Optimization (MOO). MOO can be defined as the problem of finding a vector of decision variables (actions). These variables satisfy constraints and optimize a vector function whose elements represent the objective functions. These functions form a mathematical description of performance criteria. Hence, the terms “optimize” means finding such a solution, which would give the values of all the objective functions acceptable to the designer.

## B. Prospects

As in the present network, in case we try to implement the adaptivity approach, we will only achieve a sub-optimal behavior, done by the dynamic algorithm. The approach which we are trying to implement must be able to abstract and isolate high level goals from low level actions, to integrate and act on imperfect and conflicting information, and to learn from past actions to improve future performance. This is what is required for making the Internet’s environment work efficiently, have a competing objective, decentralized control, complexity and adapt to dynamic changes.

This significant challenge is shown by the knowledge plane as described in [8]. It involves functioning well in the presence of incomplete, inconsistent, and possibly misleading or malicious information. The forces conspiring this are system failures, information filtering for privacy or competitive reasons and finite network resources to list a few. It must perform appropriately in the presence of conflicting or inconsistent higher-level goals among the Internet’s different stakeholders. This is a manifestation of the *tussle* dilemma. It must operate effectively when a new technology is introduced, or when a new application is conceived during the working of an already started application, which cannot be considered during the start of the design as in a dynamic environment where both short-term and long-term changes in the structure and complexity of the underlying network [9] are of concern.

## C. Applications

Cognitive networks are more closely related and work with wireless solutions. As studies have shown cognitive radio involves a simple model of cognition and learning [32]. Other wireless solutions predict that the networks, network components, as well as networked devices and applications, can be deployed and managed (configured, optimized, healed and protected), in real-time. Cognition can be used to improve the performance of resource management, quality of service (QoS), security, access control, or many other network goals. Cognitive networks are only limited in application by the adaptability of the underlying network elements and the flexibility of the cognitive process. The cost (in terms of overhead, architecture, and operation) must justify the performance. Thus, in almost all cases, implementing a cognitive network requires a system that is more complex than a non cognitive network [5].

In the definition of the cognitive networks, the critical definitions are network and the end-to-end aspects. For the network, the ability to self-adapt and self-organize in a changing environment has become a key issue. Since the main control function like the above self-control mechanisms, is shifted to end-user nodes. It must be equipped with mechanisms permitting them to adapt to the current network status. For this reason, biologically inspired approaches seem promising since they are highly capable of self-adaptation, although they can be slow to adapt to environmental changes [6].

The central theme of evaluation and analysis of cognitive networks is based on a three-design decision: selfishness, ignorance and control. A metric called the “price of a feature” [7] is specifically developed to evaluate the impact of these decisions on the network goals. This metric is also used to measure the expected and bounded performance of the network under the influence of each design decision.

The following section deals with the prospects of a cognitive system, the related work in the area and current software used for development. It concludes with ongoing work and future work in cognitive networks.

## III. Related Work

In the following section two major applications that incorporate a few aspects of cognitive networks have been given in detail.

### A. Cognitive Radio

Cognitive Radio is an intelligent wireless communication system that is aware of its surrounding environment and grows by learning from its environment. It adopts internal states to the incoming RF stimuli by making corresponding changes in certain operating parameters (e.g., transmit-power, carrier-frequency, and modulation strategy) in real-time, with two main purposes:

- Highly reliable communications whenever and wherever needed.
- Efficient utilization of the radio spectrum. [10, 11]

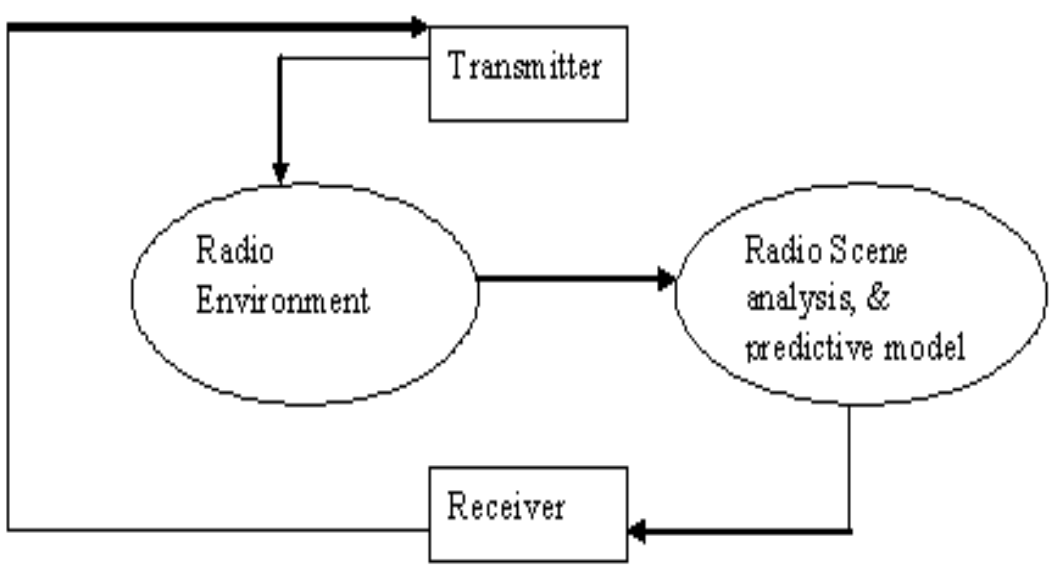

**Figure 3** Cognitive cycle in radio environment

We can see that 50%of the nomenclature in Cognitive radio has led to further development in cognitive networks. The most important thing is to learn from past decisions and use them for future behavior. Both are goal-driven and depend on observations and knowledge of node to attain decisions. Knowledge in cognitive radio is represented by RKRL (Radio Knowledge Representation Language). The two attributes of cognitive radio are goal oriented and achieve context-awareness. For this to exist in cognitive network, a network level equivalent is a must. The optimization space requires tunable parameters in cognitive radio. These are provided by SDR (Software Defined Radio). This is analog to SAN (Software Adaptable Network). Therefore, both technologies employ a software tunable platform that is controlled by the cognitive process.

### A. (a) Difference between cognitive network and cognitive radio

As compared to cognitive radio, cognitive network has more scope in controlling goals *i.e.*, a cognitive network’s goal is end-to-end and a cognitive radio is localized only for that radio’s user. These end-to-end goals are derived across the network from operators, users, applications, and resource requirements. It also helps operate more easily across all layers of the protocol stack. Currently, the work under cognitive radio is limited to the physical layer, which limits the direct impact of changes made by the cognitive process to the radio itself and other radios to which it is directly linked. Agreement with other radio for optimization must match with parameters for a successful link communication. To include all nodes impacted by the change, cognitive radio must see further physical layer. However, as the negotiation process is unable to assign precedence to radios’ desires without goals of a broader scope, achieving agreement among multiple nodes may be a slow process. This may lead to a sub-optimal network performance. In contrast, cognitive networks are more cooperative in nature, since the performance is referenced to end-to-end goals and nodes within a single cognitive element and it must cooperate to enact decisions. Another important difference lies in the heterogeneity [36] that is supported. Cognitive radio is used only by wireless network, whereas cognitive network is used by both wired and wireless network. Due to heterogeneity in cognitive network, it is better to optimize, as it is difficult to integrate. A good comparison may be carried out in cognitive processing in cognitive network with respect to multiple nodes against cognitive radio. A cognitive network has the option to implement a centralized cognitive process, a fully distributed cognitive process, or a partially distributed cognitive process.

### B. Cross Layer Design

A cross layer design means sharing of internal information between the layers and direct communication with non-adjacent layers, thus violating the traditional approach as shown in Figure 4 [19]. In the current layered architecture information is not available externally. This is indirectly taken care by the cognitive network. Thus cognitive network follows a cross-layerdesign approach. The commonality between the two is that, observation is made externally and a layer other than the layer making the observation makes some adaptation. In a cognitive network, protocol layers provide observations of current conditions, which are input to the cognitive process. The cognitive process then determines what is optimal to the network and configures the network elements protocol accordingly.

### C. (a) Differences between Cross-Layer Design and Cognitive Network

In spite of commonalities between the two, cognitive network has to deal with multiple goals and in order to achieve this, it performs multiple-objective optimizations whereas cross-layer design performs single objective optimization. Cross-layer design accounts for network wide performance goals. Trying to achieve each goal independently is likely to be suboptimal, and as the number of cross-layer designs within a node grows, conflicts between the independent adaptations may lead to adaptation loops [13].

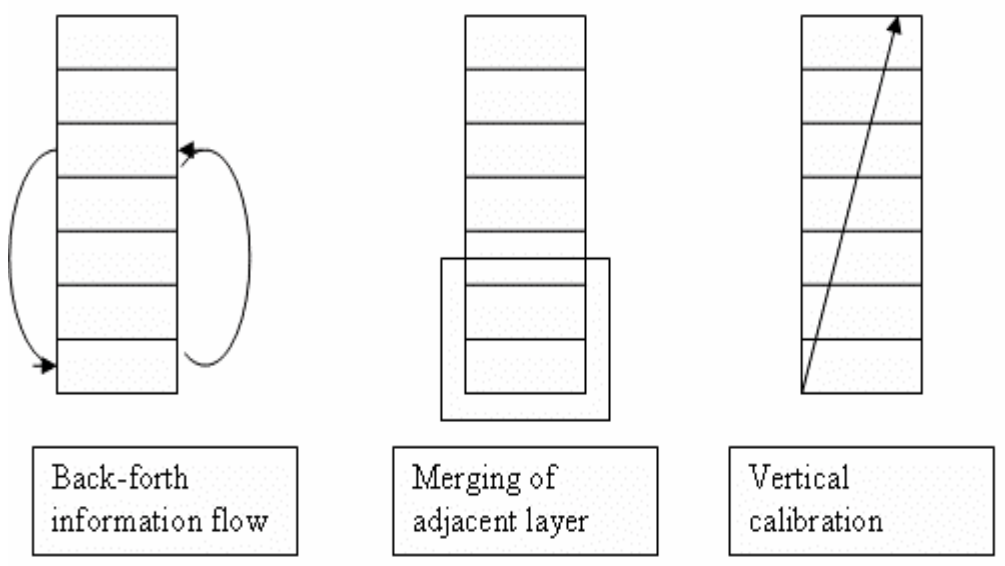

**Figure 4** Cross Layer Design

This is overcome by a cognitive network by considering all goals together in the optimization process. The other difference lies in learning [14]. The cognitive network learns from prior decisions and applies the learning to future decisions. Cross-layer designs are memory less adaptations that will respond the same way when presented with the same set of inputs, regardless of how poorly the adaptation may have performed in the past. Since our understanding is limited in the interaction between the layers, learning plays an important role. The scope of the goals and observation sets are different in both. That is, the observations used by the cognitive process span multiple nodes and the optimization is performed with the goals of all nodes in mind whereas cross-layer design is node-centric. This global information makes cognitive process to adapt more easily as compared to cross layer design.

## IV. Preliminary Area

### A. Focus of Interest

Cognitive processes fall under the title of 'machine learning' and have been an active research area since the 1960s. Machine learning is broadly defined in [15] as any algorithm that "improves its performance through experience gained over a period of time without complete information about the environment in which it operates." This definition further widens the scope of study to artificial intelligence, decision-making and adaptive algorithm, which helps in the learning process. Machine learning aims to understand computational mechanisms. This experience can lead to improved performance. In everyday language, we say that a person has 'learned' something from an experience when he can do something he could not, or could not do as well, before that experience. The field of machine learning attempts to characterize how such changes can occur by designing, implementing, running and analyzing algorithms that can be run on computers. This gives a wider approach to statistics, cognitive psychology, information theory, logic, complexity theory and operations research, with the goals of computational character of learning [16].

For cognitive networks, a neural network can be used, as it uses bottom-up method for learning, simulating the biological neurons and pathways that the brain is believed to use. These artificial neurons analyze different aspects of known inputs with some amount of unknown corruption. If network responses are modeled as a noisy pattern, a neural network could be used to categorize the pattern into predetermined responses. Kalman [17] filters contain an adaptive algorithm for feedback control. It can be used in systems that contain noise. The Kalman filter is a recursive filter used to estimate the actual and future state of the system based on noisy Gaussian measurements. It has the ability to act dynamically and is useful for tracking and maintaining a particular performance metric in a changing or noisy system. Learning automata are a distributed, adaptive control solution to identify the characteristics of an unknown feedback system. Learning automata maintains a probabilistic function for deciding what action to make. The function converges to decisions that generate desired responses in the system.

Battery-powered devices create challenging problems in the case of routing in wireless mesh networks, in terms of prolonging the autonomous lifetime of the network. For designing intelligent routing protocols, in multi-hop wireless networks, such as power-limited sensor networks, it may lead to a set of optimization problems in routing path length, load balancing, consistent link management and aggregation. For this, we can use *Supervised learning* which is a general method fortraining a parameterized function approximator, such as a neural network, to represent functions. However, supervised learning requires sample input-output pairs from the function to be learned. In other words, supervised learning requires a set of questions with the right answers. Reinforcement Learning combines the fields of dynamic programming and supervised learning to yield powerful machine-learning systems. Reinforcement learning is used popularly because of its generality. In Reinforcement Learning, the computer is simply given a goal to achieve. The computer then learns how to achieve that goal bytrial-and-error interactions with its environment. Depending on the speed of the Mobile Nodes (MNs), mobility can generally be classified into three categories of increasing speed: static, low mobility, and high mobility. The management of a network should be able to take into account any of these three cases and their associated performance implications. In the case of low mobility, the steady-state performance should be optimized and incidental updates (e.g., for route discovery) can be allowed to consume more resources; whereas in the high mobility case, resource consumption and delay due to route maintenance and updating are more important factors.

Reinforcement Learning is a form of Machine Learning, characterized by the formulation of policy to achieve specific goals. Reinforcement Learning problems are typically modeled by means of *Markov Decision Processes* (MDPs) [18].

In the above described routing approaches, reinforcement learning, Q-Learning in particular, is used to determine the actual routing paths in the multi-hop wireless network. This solution is productive if one is merely concerned with the cross-layer design of a specific routing protocol. However, it is quite clear from the literature that no single routing protocol will facilitate robust performance in all scenarios [19]. Mobile and static wireless mesh networks have fundamentally different needs in terms of route management and routing traffic generation, as well as the support of application-layer QoS needs. When concerned with the creation of a general mesh management framework, which caters to the needs of a dynamic network, a far better approach is to have access to a dynamic proactive and reactive routing protocol for relatively static and mobile mesh networks respectively. Depending on network characteristics, the appropriate protocol can be implemented and augmented in real-time by the intelligent network management agent(s) responsible for self-management and reconfiguration.

### B. Software Adaptation

For changing user needs and environment, it is better for the software to change its structure and pattern of execution dynamically. How, when and where recomposition occurs [20], is referred to as compositional adaptation. Motorola is a pioneer in the domain of cognitive networks. Its thought leadership has been captured within the End-to-End Reconfigurability (E2R) project; a Motorola led collaborative initiative comprised of 27 top academics, equipment suppliers, network operators and regulatory policy makers. Cognitive network teamwork has significant potential in the development of new public-safety, entertainment and military applications [19], experience with cognitive architectures and recent work in multiagent. More recently, the End-to-End Reconfigurability Project II (E2R II) [21, 34], m@ANGEL platform [22], CTVR at Trinity College [23], and the Institute for Wireless Networks at RWTH Aachen University [24] have proposed architectures at various degrees of maturity for end-to-end oriented, autonomous networks. Systems and the emerging field of algorithmic game theory may prove directly useful. It can be applied to many problem-solver techniques which use creative thinking [25].

## V. Challenges

The technical challenges, which have been encountered in cognitive networks are reliability and rapid sensing of the channel allocation [26, 27]. Optimization decision, learning, reconfigurable radio and security are the other major challenges. There is no fully open spectrum available and no published standard for a common platform. Legacy supports like Legacy Radios in Cognitive network and Cognitive Radios in legacy network need to be worked on. It is a NP-hard problem with existing sub-optimized solutions. Cooperation of inter-system communication, overhead and delay need to be looked at to strike a balance between complexity and performance. Mobile Ad-hoc Network Interoperability and Cooperation (MANIAC) is another challenge with respect to the present network.

## VI. Other Approaches

Among the people who are working in this area, Mitola identifies six processes which together allow a cognitive system to *'employ model based reasoning to achieve a specified level of competence in radio-related domains'* [28]. These processes are:

1. Observing the outside world.
2. Orientation of the system.
3. Planning one or more courses of action.
4. Deciding upon a course of action.
5. Acting to influence the operation of the system.
6. Learning from experience.

The process of *observing* the outside world involves the acquisition of knowledge by a cognitive radio about its environment through the analysis of incoming information streams. The *orientation* process of the cycle is concerned with establishing the priority attached to any observations made. Certain observations may require immediate action while others may feed into other cycle processes such as the *planning* process. This process is responsible for generating and analyzing courses of action. At the *decision* stage of the cycle, one of the selected courses is chosen. Finally, the decision is put into *action* and the operation of the cognitive radio is actually influenced. The *learning* process allows the system to learn from experience. Haykin [11] and Thomas [30] have similar cycles to describe the operation of cognitive radio.

Sutton [23] has given another reformation of cognitive network process by which a simplified reproduction of the cognition cycle can be drawn as shown in Figure 5. In this, the cognition cycle can be divided into two entities as illustrated. The first entity, which is termed the Cognitive Engine, broadly concerns itself with reasoning, cognition and deduction. The second entity is focused solely on the processes of observation and action. This is the *Reconfigurable Node*. The Reconfigurable Node forms a platform for cognitive networks by providing an architecture designed specifically for reconfiguration and observation, not only at the radio layers of the individual node, but also throughout the node network stack and on a network-wide scale.

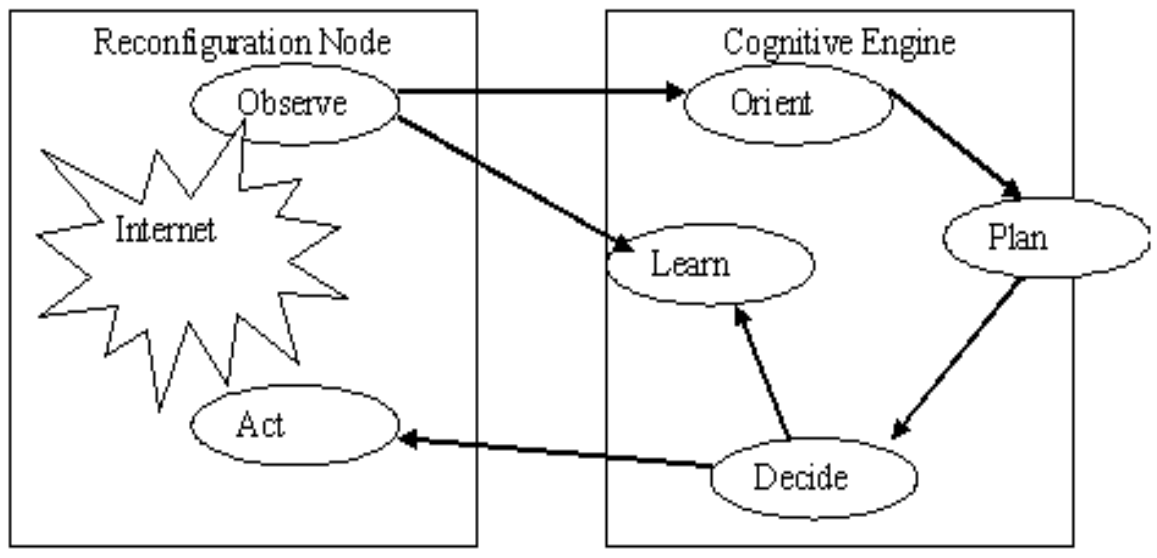

**Figure 5** Reconfigurable node

With all these developments in this area, one needs to take care of three critical design decisions that affect the performance of cognitive networks: the selfishness of the cognitive elements [33], their degree of ignorance, and the amount of control they have over the network. We define a metric called the price of a feature [31], which takes care of the three decisions, defined as the ratio of the network performance with a certain design decision to the performance without the feature. To further aid in the design of cognitive networks, we identify classes of cognitive networks that are structurally similar to one another. The two classes are: the potential class and the quasi-concave class. Both classes of networks will converge to Nash Equilibrium under selfish behavior and in the quasi-concave class this equilibrium is both Pareto and globally optimal. There is a need to identify the problems in self-organizing wireless networks that fall under these classes.

## VII. Future Work

Future work in the areas of opportunistic spectrum access, cross-layer optimization, reconfigurable protocol layers and policy-based reasoning in cognitive networks can be the major situation to be addressed [23] and designed.

## VIII. Conclusion

Thus, the future of the Internet lies in making the network intelligent. Much work has been in progress and developed for better understanding of the cognitive process. This has led to a different dimension in the growth of communication network. Thus, cognitive network area is left with many open-ended questions like prices of anarchy, ignorance and control, which exist in any design. The cognitive network has to limit its observations as much as possible just to make cognitive processing for a network feasible. Thus, this paper summarizes the current trends and approaches that have been carried out till date.